\documentclass[useAMS,usenatbib]{mn2e}
\usepackage{epsfig}
\title[Disky \& Boxy Ellipticals]
{On the Origin of Isophotal Shapes in Elliptical Galaxies}
\author[S. Khochfar \& A. Burkert]
  {S.~Khochfar,$^{1,3}$\thanks{sadeghk@astro.ox.ac.uk}
  A.~Burkert,$^{2,3}$ \\
  $^1$ Department of Physics, Denys Wilkinson Building, Keble Road, Oxford OX1 3RH, United 
Kingdom \\
$^2$ University Observatory Munich, Schreinerstr. 1, 81679 Munich, Germany \\
 $^3$ Max-Planck-Institute for Astronomy, K\"onigstuhl 17,
69117 Heidelberg, Germany\\
   }
\date{Released 2004 Xxxxx XX}

\pagerange{\pageref{firstpage}--\pageref{lastpage}} \pubyear{2004}

\def\LaTeX{L\kern-.36em\raise.3ex\hbox{a}\kern-.15em
    T\kern-.1667em\lower.7ex\hbox{E}\kern-.125emX}

\begin{document}

\label{firstpage}

\maketitle

\begin{abstract}
Using semi-analytical models of galaxy formation, the origin of
boxy and disky elliptical galaxies is investigated. We find that the simple scenario,
motivated by N-body simulations, in which the isophotal shape is only dependent
on the mass ratio of the last major merger, is not able to reproduce the observation
that the fraction of boxy and disky ellipticals depends on galaxy luminosity. The
observations can however be reproduced with the following reasonable assumptions:
(i) equal-mass mergers lead to boxy ellipticals and unequal-mass mergers produce disky
ellipticals (as motivated by N-body simulations) (ii) major mergers between bulge-dominated
galaxies result always in boxy ellipticals, independent of the mass ratio,
(iii) merger remnants that subsequently accrete gas leading
to a secondary stellar disk with more than $20 \%$ the total stellar fraction are always
disky. This scenario indicates that the isophotal shapes of merger remnants
are sensitive to the morphology of their progenitors and subsequent gas infall.
Boxy and disky ellipticals can be divided into two subclasses, depending on their
formation history. Boxy ellipticals are either formed by equal mass mergers  of disk galaxies
or by major mergers of early-type galaxies. We find that
disky ellipticals are indicators of unequal-mass mergers or late gas infall.
Disky ellipticals with high luminosities are preferentially  1-component systems that 
result from unequal mass mergers whereas low-luminosity disky ellipticals are more likely
to harbour secondary disk components. In addition,
the fraction of disky ellipticals with  secular disk components should increase in 
regions with higher galaxy densities. Taking into account the conversion of cuspy cores
into flat low-density cores by black hole merging we find that disky ellipticals should 
contain central density cusps whereas boxy ellipticals should in general be characterised
by flat cores. Only rare low-luminosity boxy ellipticals, resulting from equal-mass mergers 
of disk galaxies could have power-law cores.
\end{abstract}


\section{Introduction}

Elliptical galaxies are old stellar systems that are believed to have
been formed by major mergers of disk galaxies, preferentially
at high redshift \citep{tt72,s73,c04}.
This ``merger hypothesis``  has been tested using numerical simulations
 \citep[e.g.][]{g81,n83,b88,h92} which
demonstrated consistently that the global properties of
equal mass merger remnants resemble those of ordinary slowly rotating
massive elliptical galaxies.

More recently it has become clear that ellipticals are more complex than originally thought.
Isophotal fine structures have been detected that
correlate well with kinematical properties \citep{ben88}.
Faint ellipticals are isotropic, fast rotators with small minor
axis rotation. They are called disky as
a Fourier analysis of their isophotal deviations from perfect ellipses 
leads to positive values of
the fourth order coefficient $a_4$. Disky ellipticals might contain secondary,
faint disk components which contribute up to 30\% to the total light in the galaxy, indicating
disk-to-bulge ratios that overlap with those of S0-galaxies \citep{rw90,sb95}.
In contrast to disky ellipticals, boxy systems are characterised by
negative values of $a_4$. They rotate slowly and are supported by velocity anisotropy.
Their cores are flat \citep{lau95,fab97} and reveal complex internal kinematics.
The distinct physical properties of disky and boxy elliptical galaxies
indicate that both types of ellipticals experienced different
formation histories. Interestingly, most of the massive ellipticals are boxy, while
2/3rd of the lower-mass ellipticals are disky \citep{ben92}.

In order to understand the origin of boxy and disky ellipticals within the framework
of the major merger scenario, numerical merger remnants of dissipationless mergers 
have been analysed in detail e.g. by \citet{h93}, \citet{hey94}, \citet{ln95},
\citet{nbh99} and \citet{bendo00}. In addition the role of gaseous dissipation 
on the the isophotal shape of remnants has been investigated in various studies e.g.
by  \citet{bah96}, \citet{bs97} and \citet{spr00}. These studies find in agreement 
with each other that gaseous dissipation leads to enhanced diskyness in the remnant.
 \citet{bs97} simulated a number of mergers between bulgeless disk galaxies with 
purely gaseous disks which depending on the rapidity of star formation tend to be
more likely disky (boxy) for weak (strong) star formation efficiency. This result
can be understood in terms of gas that has lost large parts of its angular momentum 
and settles down in the centre of the remnant, increasing the potential well and 
thereby destabilising box-orbits passing the central region  \citep{bah96}. The amount 
of gas able to settle down into the centre is strongly correlated with the efficiency 
by which gas is transformed into stars which explains the trend observed in the simulations.
However, \citet{spr00} demonstrated that this effect is much weaker in the presence of a 
stellar bulge component. \citet{kh00} and \citet{kb03} showed that the role of 
dissipation in the formation of massive spheroids is weaker compared to its role 
in the formation of low mass spheroids. Furthermore massive spheroids tend to form 
mainly by mergers of progenitors already consisting of massive bulge components \citep{kb03}.
In the following, we will compare our results to the giant and parts of the intermediate 
elliptical sample  with absolute B-band magnitude $M_B \leq -18.5$ of \citet{ben92}
allowing us to neglect the influence of dissipation during the merger process 
on the isophotal shape.

Recently, \citet{nb03} \citep[for a summary see also][]{bn03} completed a large survey of
dissipationless merger simulations of disk galaxies, adopting a statistically 
unbiased sample of orbital initial conditions with mass ratios of 1:1 to 4:1.
They showed that unequal mass 3:1 to 4:1 mergers lead to fast rotating
disky ellipticals, in excellent agreement with observations. In contrast, slowly rotating, 
pressure supported ellipticals formed in equal mass 1:1 to 2:1
mergers of disk galaxies while 2:1 to 3:1 mergers lead to remnants with mixed isophotal shapes.

In this paper we investigate whether the scenario of disky and boxy elliptical formation 
based on mass ratios in major mergers is in agreement with the observed distribution of 
ellipticals with different isophotal shape.
Using semi-analytical simulations we predict the ratio of boxy and disky ellipticals as a
function of their luminosity. In section 2 \& 3 we introduce the semi-analytic approach
 we use and the way in which we assign ispohotes to modelled ellipticals. Section 4 
compares our model predictions with observations and in section 5 we introduce two 
new sub-classes of elliptical galaxies followed by a concluding section.

\section{Formation of Ellipticals}

We follow the formation and evolution of elliptical galaxies in the 
context of semi-analytical modelling of galaxy formation. The history of 
dark matter halos is traced using the merger tree proposed by 
\citet{som99} and the baryonic physics is treated as described e.g. in 
\citet{spr01}. The dark matter merger trees are resolved down to a minimum 
mass of $M_{min}=10^{10}$ M$_{\odot}$ and calculated for a set of different present 
day dark halo masses ($M_{0}=10^{13}$, $10^{14}$,
 $10^{15}$ M$_{\odot}$). Larger halo masses correspond to more dense environments, as 
e.g. $ M_{0}=10^{15}$ M$_{\odot}$  and $ M_{0}=10^{13}$ M$_{\odot}$ 
corresponds to a present day cluster and field environment, respectively. 
The resolution limit we adopt does not 
influence our results as we will show in the next section. We assume a 
$\Lambda$CDM  model with cosmological parameters $\Omega_m=0.3$, 
$\Omega_{\Lambda}=0.7$, $h=0.7$, $\Omega_m/\Omega_b=0.16$ and $\sigma_8=0.9$.

The galaxy formation paradigm inherent to semi-analytic 
models assumes that the first galaxies to form are bulge less disk galaxies 
\citep[see however][]{db04} which form stars in centrifugally supported disks of cooled gas 
\citep[e.g.][]{wf91}. 
Subsequently spheroids, elliptical galaxies and bulges,
 are formed in major mergers with mass ratios $M_{gal,1}/M_{gal,2} \leq 3.5$ ($M_{gal,1} 
\geq M_{gal,2}$) of galaxies as proposed by \citet{tt72}
 \citep[e.g.][]{k99,spr01}.  We adopt a scheme identical to the one 
used in e.g. \citet{spr01} in which
each galaxy in the semi-analytical simulation consists of two stellar components, a disk and 
a bulge component. 
\subsection{Disk Component}
The disk component grows through hot gas that radiatively
cools from the halo region and settles into the equatorial plane of a galaxy
where it is transformed into a secular stellar disk. The cooling rate 
is calculated using the prescription
described in \citet{spr01}. 
Additionally we allow disks to grow by accretion of the cold gas in 
the disks of satellite galaxies in minor mergers. We neglect any further possible 
ways of growing disks and comment on that in the next section. 
\subsection{Bulge Component}
The bulge component grows through major mergers of galaxies. We calculate the 
timescale for galaxies to merge using the approach in \citet{k99} with the 
modification of using the approximation of the Coulomb logarithm as proposed 
in \citet{spr01}. Mergers disrupt the progenitor disks as seen in various 
numerical simulations  
\citep[e.g.][and reference therein]{ba92,bn03} and relax to a spheroidal 
distribution. During the merger any cold gas in the disk of the progenitor galaxies is
assumed to be funnelled into the centre of the remnant where it
ignites a star burst which transforms all of the cold gas into stars contributing to the 
spheroidal component \citep[e.g.][and reference therein]{k99,spr01}. The second 
assumption is certainly a 
simplification of what might happen since we neglect that not all of the cold gas is 
funnelled to the centre but some fraction of it can e.g. settle down in an extended disk 
 which continues growing inside out by fresh supply of gas from tidal 
tails \citep[e.g.][]{bh91,mih96,ba01,b02}. \citet{b02} results indicate that $40\% - 80\%$ 
of the initial gas could end up in the central region of the remnant and be consumed 
in a star burst. The exact amount is somewhat dependent on the merger geometry and 
on the mass ratio of the merger. Unfortunately, a large survey  
investigating the gas inflow to the centre of merger remnants is not available to date 
so that we use the simplified approach.
The prescription for the faith of 
the cold gas we adopt results in an overestimate of the spheroid masses and an underestimate 
of the secondary disk components in our model, which is not very significant  
for massive ellipticals since they are mainly formed in relatively weak dissipative 
mergers \citep{kh00,kb03}. Another simplifying assumption is that we neglect the 
 feeding of super massive black holes in the centre of the remnant or feedback 
effects on the gas from the central source. However, \citet{hk00} estimate that a 
cold gas mass fraction of less than 1\% accreted onto the black hole is sufficient 
to recover the $M_{\bullet}-\sigma$ relation and we therefore neglect this effect. 
 Furthermore we assume, that the bulge components of galaxies grow in minor mergers by 
the stars of infalling satellites.
\subsection{Morphological Classification}
The above description for the growth of different stellar components allows 
for morphological transitions of individual galaxies. All galaxies start off as bulgeless 
disk galaxies which later transform into elliptical galaxies in major mergers, 
and  then possibly grow
new secondary disks through the combined effect of cooling of hot gas from halo regions 
and cold gas accretion in minor mergers. 
In this way, provided that enough cold gas is accreted, the ellipticals could also become bulges of 
early-type spirals until the next major merger which destroys the disk again and creates 
a new elliptical galaxy when the  process of disk growing starts again.
The knowledge of the bulge and disk components allows a morphological classification 
of modelled present day galaxies using the correlation between the Hubble type $T$  
and the $B-$band bulge-to-disk ratio of galaxies presented in \citet{sd86}. 
Elliptical galaxies are classified by $T < -2.5$ and have $ \Delta M=M_{B,bulge}-M_{B,tot} 
\leq 0.55 $. Our definition of morphology bares some danger with respect to 
lenticular galaxies. Generally semi-analytic models assume $ -2.5 < T < 0.92$ 
for S0 galaxies, 
identifying them as a transition type between ellipticals and spirals, in the process of 
regrowing a secondary disk after a major merger \citep{k99}. Alternative formation mechanisms 
e.g. involve either the unequal mass mergers of two spiral 
galaxies \citep{bek98,ba96,bendo00} in which
the larger progenitor's disk basically survives the merger, or the growing of a secondary 
disk by material from tidal tails \citep{mi04}. As a consequence, our population 
of ellipticals formed in mergers should be polluted with S0 galaxies. At this stage,
a clear separation between modelled disky ellipticals and S0 galaxies is not possible and  
is hindered by observational ambiguities due to projection effects. We therefore do not distinguish
between disky ellipticals and S0 galaxies for $T < -2.5$.

\section{Assigning Isophotes}

Following the work of \citet{ba98}, \citet{nbh99} and \citet{nb03} we assign isophotes to 
elliptical galaxies according to the mass ratio of their last major merger.
Last major mergers with mass ratio $ 1 \leq M_{gal,1}/M_{gal,2} < 2$ and 
$ 2 \leq M_{gal,1}/M_{gal,2} < 3.5$ $(M_{gal,1} \geq M_{gal,2})$ lead to 
boxy and disky ellipticals, respectively.

\begin{figure}
\begin{center}
\includegraphics[width=8cm,angle=0]{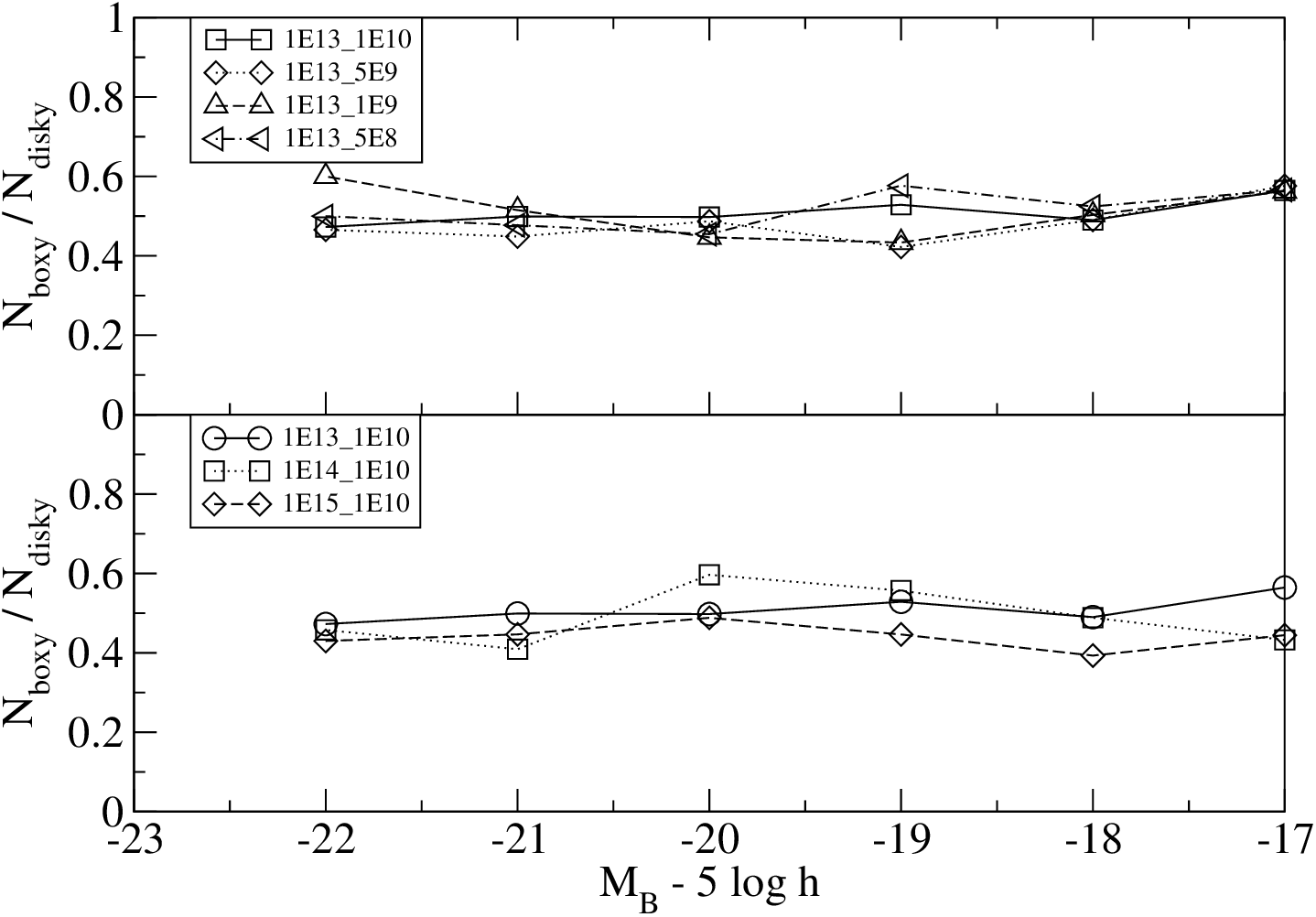} 
\medskip
\caption{Upper panel: Ratio of boxy to disky ellipticals for different 
magnitudes. Symbols refer to runs with 
different mass resolution. The labels in the graph indicate the halo mass  
at a redshift of $z=0$ (in this case $M_{0}=10^{13}$ M$_{\odot}$) followed by 
the minimum mass ($M_{min}=10^{10}$, $5 \times 10^{9}$, $10^{9}$, $5 \times 10^{8}$
M$_{\odot}$). Lower panel: Ratio of boxy to disky ellipticals for different 
magnitudes found in the simulation. Symbols refer to runs with 
different dark halo masses at $z=0$. The labels in the graph indicate the 
halo mass  at a redshift of $z=0$ (in this case $M_{0}=10^{13}$, $10^{14}$,
 $10^{15}$ M$_{\odot}$) followed by 
the mass resolution ($M_{min}=10^{10}$ M$_{\odot}$). \label{fig1}}  
\end{center}
\end{figure}

 Fig. \ref{fig1} shows the ratio of boxy to disky elliptical galaxies depending
 on their present day $B-$band magnitude, assuming a final dark halo mass of
$10^{13} M_{\odot}$. The ratio is almost constant with
$N_{boxy}/N_{disky} \sim 0.5$ over the presented magnitude range. This result 
is a consequence of the self-similar build up of elliptical galaxies by major
 mergers on all mass scales. We mentioned
 above that we follow dark matter histories back until progenitor halos drop
 below $M_{min}=10^{10}$ M$_{\odot}$. The different symbols in the upper panel
 Fig. \ref{fig1} 
show the results of a resolution study of this minimum mass. By increasing 
the resolution (i.e. decreasing $M_{min}$) we also increase the number of resolved small 
mass satellite galaxies in the overall galaxy population. As shown, the 
results do not change which is due to small mass satellite galaxies not significantly taking
part  in the last major major mergers at these luminosity scales. In the 
following we will use a mass resolution of $M_{min}=10^{10}$ M$_{\odot}$
 which is a good compromise between efficiency and accuracy.

It is well known that, observationally, the fraction of elliptical galaxies 
increases with the galaxy density of the environment \citep[e.g.][]{d80}, a trend 
also seen in simulations of galaxy formation \citep[e.g.]
[and reference therein]{spr01}, which is attributed to the increased merger rate. 
The fraction of major mergers and its redshift
 evolution are very sensitive to environmental effects \citep{kb01}. However, 
one is not expecting to find the relative fraction of major mergers for a given 
mass ratio to change with the environment. The lower panel of Fig. 1 shows the dependence of the
ratio of boxy-to-disky ellipticals
on the final dark halo mass which corresponds to different environments. Large final
masses represent higher density environments, like clusters, low-mass represent 
a field environment. There is no significant environmental dependency, suggesting 
that the mix of boxy to disky ellipticals is universal.

\section{Comparison with Observations}
The next step is to compare the model predictions to observations by \citet{ben92} 
who presented a first systematic study of the correlation of  the isophotal shape with 
other physical properties of ellipticals. 
\citet{nbh99} and \citet{nb03} used the same reduction routines as \citet{ben92}
in their merging simulations to derive the isophotal properties of their remnants.
It is therefore safe to assume that we do not introduce any 
ambiguities in our definition of boxy and disky ellipticals by adopting their 
predicted dependence of isophotal shape on the mass ratio of the merger.

\subsection{A First Test}
\begin{figure}
\begin{center}
\includegraphics[width=8cm,angle=0]{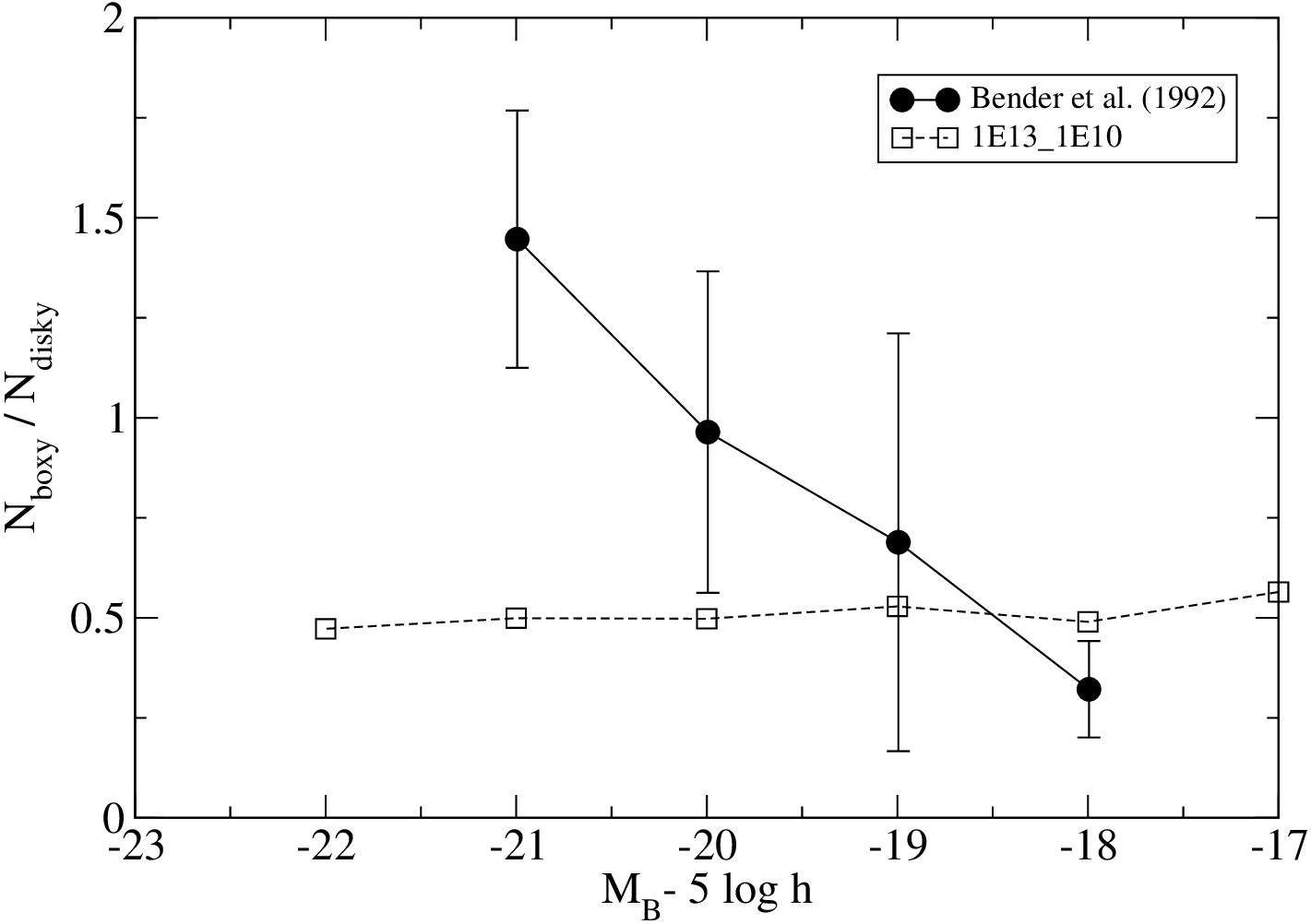} 
\medskip
\caption{Ratio of boxy to disky ellipticals for different 
magnitudes found in the simulation. Filled circles show the data of 
\citet{ben92} and filled squares the results for the run with $M_{0}=10^{13}$
M$_{\odot}$ and mass resolution $M_{min}=10^{10}$ M$_{\odot}$. \label{fig2}}  
\end{center}
\end{figure}

The data of \citet{ben92} shows a trend that the most massive ellipticals are
mainly boxy, while the least massive ones are mainly disky. We compare this data with
the model predictions in Fig. \ref{fig2}. This comparison shows a clear failure of
the theoretical model as it does not reproduce the trend seen in the observations. 
This is a 
generic feature of the cold dark matter paradigm. Any property of a galaxy 
which depends on the mass ratio of the last major merger will be scale free. 

\subsection{Modifying the Model}

The numerical simulations which motivated the previous analysis 
still leave space for further interpretation. It is e.g. not yet 
fully understood what influence gas will have on the structure of the merger
remnants. Galaxy mergers in numerical simulations
usually start with relaxed spiral galaxies even though galaxy formation models 
predict that  mergers between elliptical galaxies or elliptical and spiral 
galaxies occur too \citep{kb03}. In the following sections we therefore  study the
possible effect of mergers between different galaxy morphologies and gas infall 
on the ratio of boxy to disky ellipticals.

\subsubsection{Progenitor Galaxies}
Recent semi-analytic simulations of galaxy formation by \citet{kb03} showed a 
remarkable correlation between the luminosity of elliptical galaxies and the 
morphology of their progenitor galaxies. The most luminous elliptical galaxies 
 had preferentially experienced the last major mergers between two elliptical galaxies and not
 between two spiral galaxies. Only low luminous galaxies are formed by pure 
gas rich spiral mergers. \citet{kb96} and 
\citet{fab97} argued along the same line that the most massive elliptical 
galaxies with boxy isophotes are most probably the outcome of a dissipationless
 merger, while less massive ellipticals with disky isophotes result from the 
merger of two gas rich galaxies, which leads to the formation of a secondary disk in the 
remnant. 

The simulations analysed by \citet{ba98,nbh99,nb03} merged 
spiral galaxies. Merger of spheroids were simulated  by \citet{g93}, who showed that the 
remnant will have most likely boxy isophotes.
In the following, we will assume
 that ellipticals having last major mergers between two bulge 
dominated galaxies ($M_{bulge} \geq 0.6 M_{tot}$) will result in boxy remnants
independent of the mass ratio. For all the other major mergers we adopt the 
original prescription.

\begin{figure}[h]
\begin{center}
\includegraphics[width=8cm,angle=0]{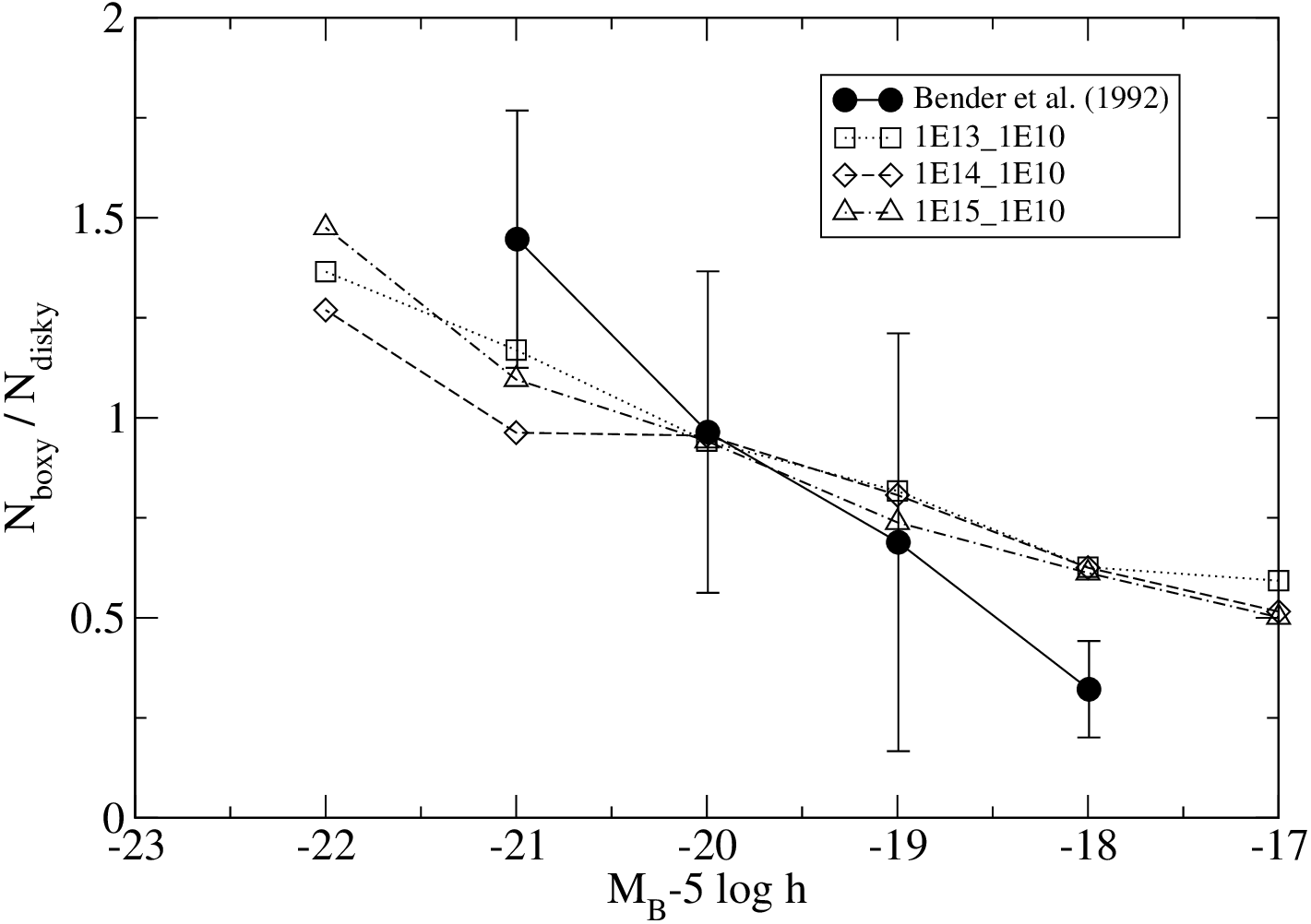} 
\medskip
\caption{The ratio of boxy to disky ellipticals as function of
 magnitudes as predicted by the simulations. Filled circles show the
observational data of \citet{ben92}, the open symbols show the 
theoretical predictions for different galaxy
environments as presented also in Fig. \ref{fig1} (lower panel). The dependence
on the morphology of the progenitor galaxies in the last major
merger is included. The mass resolution in all runs is $M_{min}=10^{10}$ 
M$_{\odot}$. \label{fig3}}  
\end{center}
\end{figure}

Figure \ref{fig3} shows the result of including this dependence on the 
progenitor morphologies for different environments. 
We now find indeed a correlation with luminosity, independent of environment.
The most luminous elliptical
 galaxies are mainly boxy and the 
fraction of boxy ellipticals decreases with decreasing luminosity. This trend 
is an imprint of the high fraction of early-type mergers for luminous galaxies.
\citet{kb03} found this fraction to be independent of environment which explains
 the results in Fig. \ref{fig3}. 
However, the gradient which we find is still not as strong as observed, although the 
observational uncertainties are large.

\subsubsection{Disks in Ellipticals}

After the last major merger, elliptical galaxies can continue to grow disks;
 one can identify three main processes of disk growth after major mergers. 
As noted above, gas in tidal tails can settle down into a disk 
after the spheroid has formed. 
We also noted above that we do not model this process and  instead assume that the gas in 
the tidal tails is accreted to the centre of the remnant at the same time as
the other present cold gas in the disks of the merging galaxies. The second and third 
process in growing 
a disk involve the cooling of hot gas from halo regions and the accretion of cold gas 
from satellite galaxies in minor mergers, respectively. 
In the following we are going to investigate 
the influence of the last two processes mentioned above on the mix 
of ellipticals with different isophotal shapes.    

\begin{figure}[h]
\begin{center}
\includegraphics[width=8cm,angle=0]{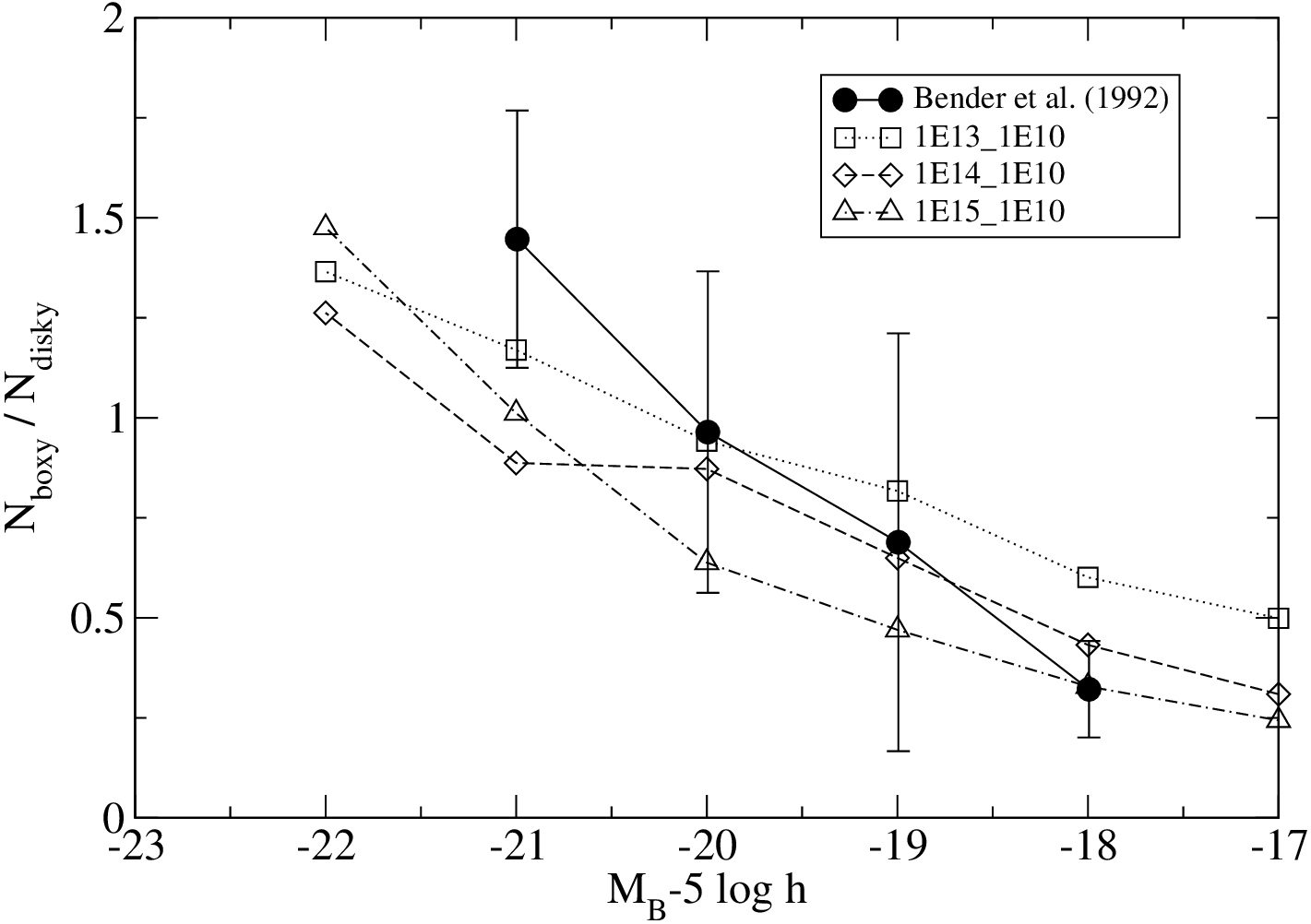} 
\medskip
\caption{Ratio of boxy to disky ellipticals for different 
magnitudes found in the simulation. Filled circles show the data of 
\citet{ben92} and open symbols the results for the environments 
presented in Fig. \ref{fig1} (lower panel) including the dependence of the 
isophotal shape on the morphology of the progenitor galaxies in the last 
major merger of ellipticals and the formation of a secondary disk by
gas infall into the 
remnant. The mass resolution in all runs is $M_{min}=10^{10}$ M$_{\odot}$. 
\label{fig4}}  
\end{center}
\end{figure}

\citet{rw90} find that some ellipticals show evidence for embedded
stellar disks with up to 30$\%$ of the total baryonic mass. 
It is likely that ellipticals with 
disks of that order will be disky, independent of the details of the last major merger.
Figure \ref{fig4} shows the results if we include this effect in our model. 
Here we assume that an elliptical will be disky if a stellar disk component is present,
 generated by the combined effects of cooling and satellite accretion, which 
contains more than $20\%$ of the total baryonic mass.
 This value is a lower limit based on estimates derived from
artificially adding  an additional stellar disk to a merger remnant and deriving the 
isophotal shape of it (Naab, private communication). We ran some tests with different values
for the lower limit when a elliptical becomes disky, finding that for a value of $30 \%$ the 
fraction of boxy to disky elliptical did not change.  
For a value lower than $20 \%$ the fraction of boxy to disky elliptical galaxies decreases,
 especially
for low-luminosity systems which on average have their last major merger earlier in contrast to
high luminosity elliptical galaxies, which do not have enough time 
after their last major merger to grow a disk that contributes significantly to the 
total light.
The fraction of ellipticals that are classified as disky because of an additional 
disk component depends also on the environment.  In high 
density environments, the last major merger occurs on average at a larger 
redshift giving a new disk more time to grow; this  explains the larger 
fraction of ellipticals with secondary disk components 
in high density environments on all magnitude scales. Note however that these results depend
on the efficiency of gas stripping in clusters (e.g. Mori \& Burkert 2000).

\section{Isophotal Sub-Classes}

The results of the last section suggest that boxy as well as disky isophotes 
have two distinct origins in ellipticals. Each formation scenario is expected to 
leave signatures in the structural properties of the remnant. Observations of 
surface brightness profiles also suggest a dichotomy  \citep{lau95,geb96,fab97}. 
Similar to the separation into boxy and disky isophotes, ellipticals can
 be classified into core and power-law galaxies, depending on the inner slope 
of the surface brightness profile. High luminous elliptical galaxies show 
shallow inner profiles (core galaxies) and low luminous ellipticals show steep
 inner slopes (power-law galaxies). It is still a matter of debate whether this dichotomy is 
real or artificial as argued by \citet{gg03} who find  a continuous 
linear relation between the Sersic index $n$ and the absolute magnitude of the
 elliptical galaxies. For our present analyses we will assume that a dichotomy exist.

\citet{re01} analysed the correlation between isophotal shape and 
the shape of the central surface brightness profile. They showed that most of 
the boxy ellipticals and disky ellipticals are classified as core and 
power-law galaxies, respectively. Note that not all of the  boxy ellipticals have a core 
and not all of the disky ellipticals are power-law galaxies, suggesting that 
each isophotal class might be sub-divided into two subclasses, depending 
on their core properties. In the following we investigate the origin of these
subclasses and test 
whether our model can account for the observation by   \citet{re01}.

\subsection{Boxy Ellipticals: flat-density cores versus power-law cores}

Our model predicts that boxy ellipticals  form by two mechanisms, equal 
mass mergers of two galaxies and major mergers ($ M_{gal,1}/M_{gal,2} 
< 3.5$, $M_{gal,1} \geq M_{gal,2}$) of two spheroidally dominated galaxies. It 
is widely accepted that spheroids harbour super-massive black holes (SMBH) 
which follow a $M_{\bullet}-\sigma$ relation \citep{fm00,geb00}. 
\citet{mm01} investigated the merger of two galaxies harbouring SMBHs and 
found that the remnant will have a light profile showing a core in the 
centre. We therefore assume that the major mergers between spheroidal 
dominated ellipticals will lead to the formation of core galaxies with 
boxy isophotes. Equal mass mergers between disk dominated galaxies harbouring 
no SMBH or SMBH with small masses will not lead to the creation of a core due to
binary black hole merging. In fact, gas infall into the centre during the merger
and subsequent star formation are likely to regenerate cusps. We 
therefore assume that these mergers generate power-law galaxies. 

\begin{figure}[h]
\begin{center}
\includegraphics[width=8cm,angle=0]{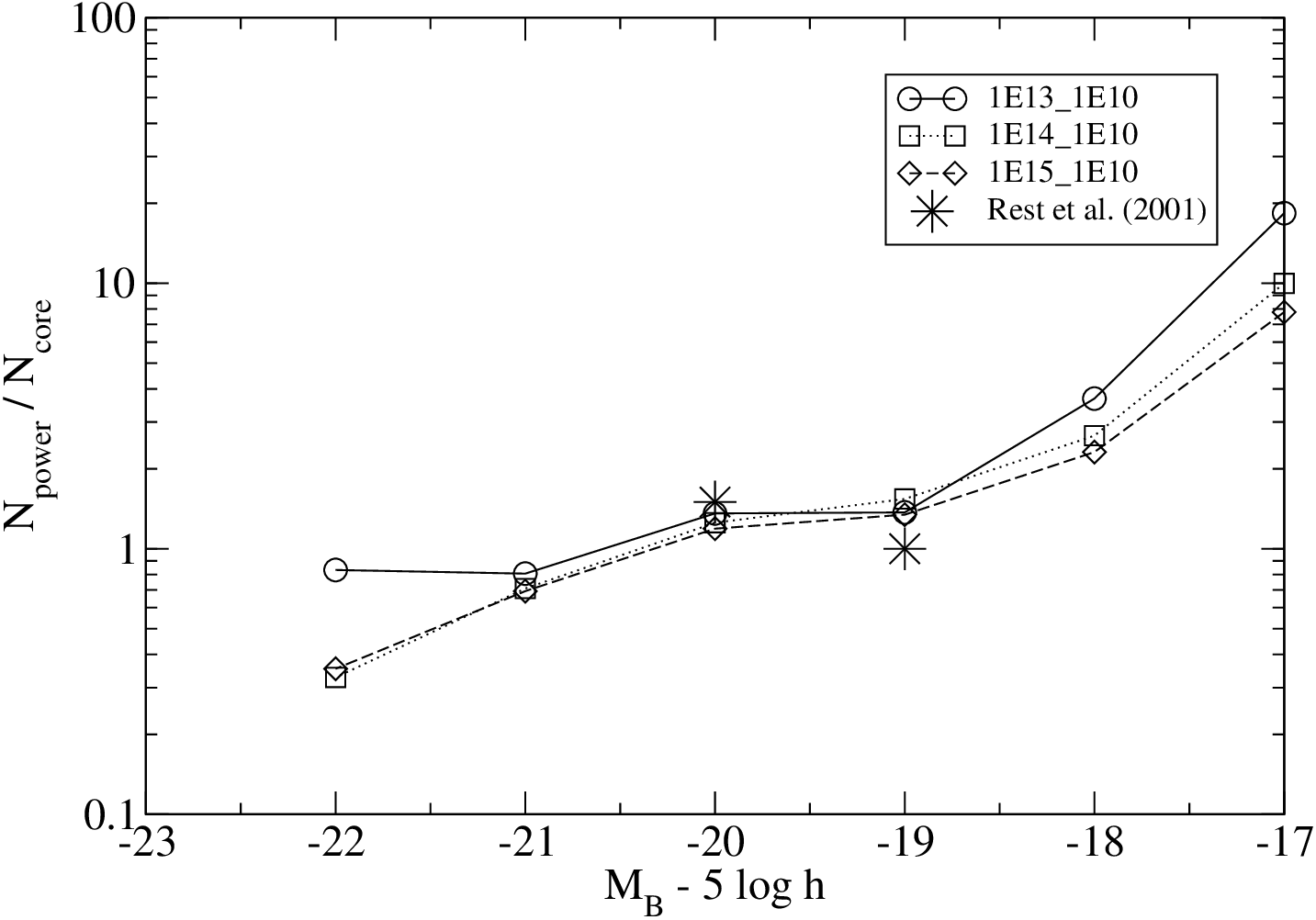} 
\medskip
\caption{Ratio of power-law to core boxy elliptical galaxies. Stars show the data 
of \citet{re01}. The other symbols show the theoretical results for the different 
environments. The mass resolution
 in all runs is $M_{min}=10^{10}$ M$_{\odot}$. \label{fig5}}  
\end{center}
\end{figure}

Fig. \ref{fig5} shows the ratio of power-law to core boxy elliptical galaxies at each magnitude
for different environments. We find no significant
environmental dependence. The
number of core galaxies increases from about 0.1 to 3 times the number 
of power-law galaxies from low to high luminosities. \citet{re01} use 
three main profile classes: core, power-law and intermediate. 
Following the original work of \citet{fab97} we classify
intermediate galaxies  as core galaxies 
and compare the data of \citet{re01} with our results in Fig. \ref{fig5}. We 
find that the agreement between the simulations and the observations in the 
overlapping luminosity range is very good. Unfortunately, due to the limited 
data, it is not possible to compare the observational data with 
the simulations over a wider range of luminosities.

\subsection{Disky ellipticals: central properties, 1-component 
versus 2-component galaxies}

\begin{figure}[h]
\begin{center}
\includegraphics[width=8cm,angle=0]{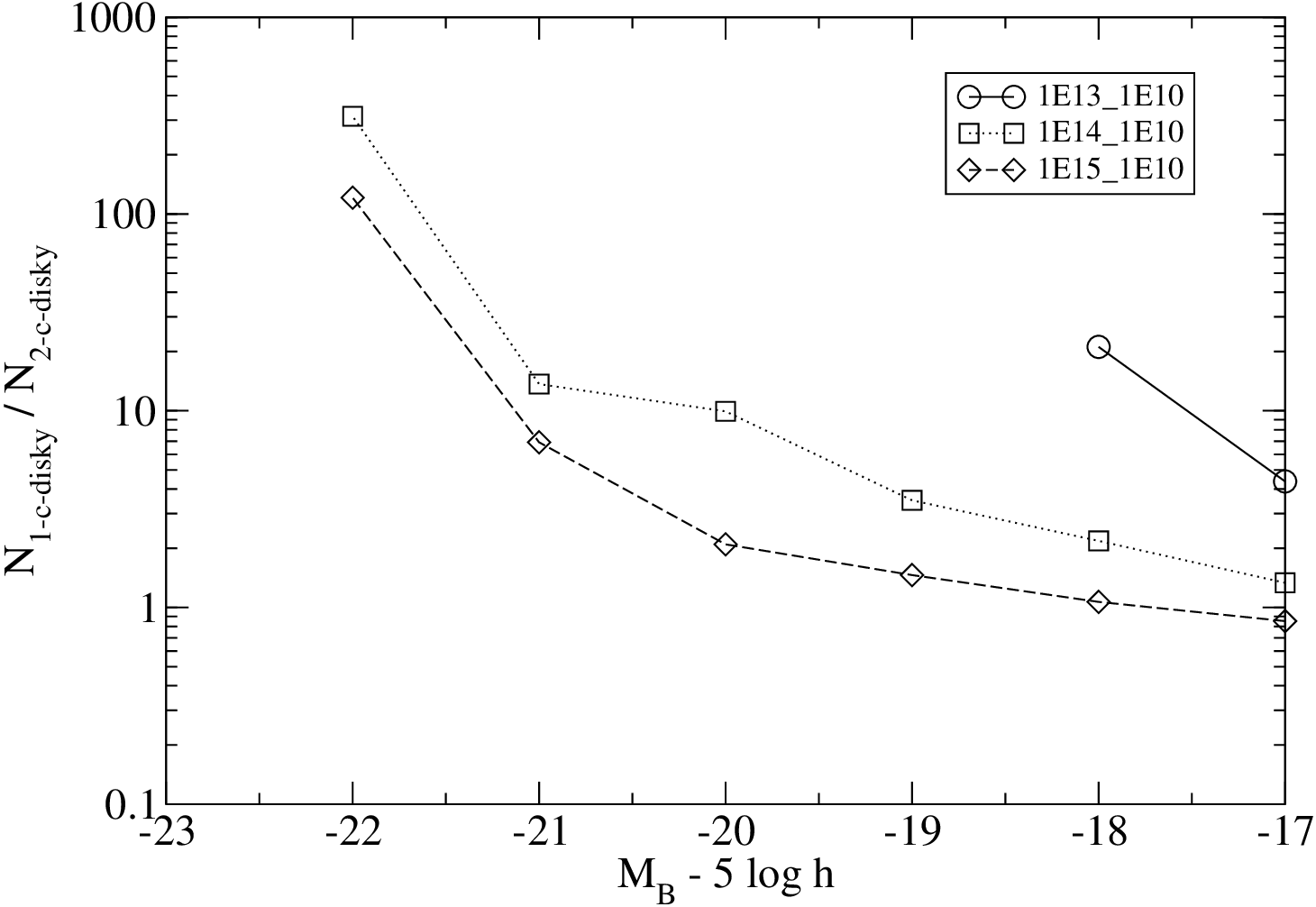} 
\medskip
\caption{Ratio of the number of disky galaxies formed in major mergers with 
$ 2 \geq M_{gal,1}/M_{gal,2} < 3.5$ ($M_{gal,1} \geq M_{gal,2}$) to disky 
galaxies which grew a disk with $M_{disk} \geq 0.2 M_{tot,baryons}$ by gas accretion. 
The symbols show the results for the different environments.
No results for magnitudes $M_{B} < -18$ in the run 
with $M_{0}=10^{13} M_{\odot}$ are shown because we did not find any disky ellipticals 
which formed due to gas accretion at these magnitudes.
The mass resolution in all runs is $M_{min}=10^{10}$ M$_{\odot}$. 
\label{fig6}}  
\end{center}
\end{figure}
 
The observations of \citet{re01} show that it is very unlikely to find 
disky ellipticals which are also core galaxies. In fact they find only 1 out of 
21. Our simple model does not allow for disky ellipticals which have a core.
The reasoning behind this is that we assume  core 
galaxies form by the mergers of two bulge dominated galaxies harbouring SMBH. This
 results in the formation of boxy ellipticals. 
In principle, it is possible to generate disky core-ellipticals; 
consider two disk galaxies with significant bulge components 
$M_{bulge} \leq 0.5 M_{tot}$. The bulges will harbour SMBH. Unequal mass mergers 
between them would result in disky ellipticals. Cores might still form when the
massive black holes spiral into the centre. At the moment, due to the lack 
of detailed simulations it is not clear whether black hole merging
in this case would be efficient enough to generate cores.
In addition, since the combined mass of the progenitor SMBHs is smaller than 
that expected for the SMBH in the remnant according to the  
$M_{\bullet}-\sigma$ relation, the black holes must grow significantly in mass 
by gas accretion in contrast to the case of two bulge dominated merging galaxies. 
Gas infall into the central region could trigger central star formation which might
change the core density distribution, leading again to cuspy profiles.
Given this complexity, we do not try to model the fraction of disky core galaxies,
which according to the observations is small anyway.

The fraction of disky elliptical galaxies with massive secular disks formed 
after the last major merger is very sensitive to both the environment and the 
amount of residual star formation in elliptical galaxies. Recent results from 
the GALEX mission indicate that there is indeed residual star formation occuring 
in elliptical galaxies (S. K. Yi, private communication). It is not yet 
clear where exactly this star formation occurs even though there are 
indications that it is in a disk-like substructure. 
Figure \ref{fig6} shows the ratio of disky ellipticals
formed by unequal mass mergers between disk galaxies 
(1-component disky) to disky ellipticals which 
have grown secular massive disks after their last major merger by gas accretion 
(2-component-disky). Most disky ellipticals are predicted to be 1-component systems
that resulted from unequal-mass mergers. 
We find a larger fraction of 2-component disky ellipticals in high density 
environments rather than in low density environments. In addition the
fraction of 2-component disky ellipticals decreases toward higher luminosities.
In high density environments, at luminosities of
$M_{B} \sim -18$, the fraction of 2-component disky ellipticals is comparable to the 
fraction of 1-component disky ellipticals.

\section{Discussion and Conclusion}

We used semi-analytical simulations to investigate the formation of disky 
and boxy ellipticals. A comparison with observations shows that 
the isophotal shape cannot only
depend  on the mass ratio of the last major merger. In this case,
galaxy properties which depend only on the mass ratio of the last major merger  
should be scale free. The fraction of
disky-to-boxy ellipticals, however, shows a strong mass dependence.
To break the mass degeneracy, we propose that the isophotal shape should, in addition to the 
mass ratio, depend also on the morphology of the progenitors of the last major merger.
Our model can reproduce well the observed trend of the ratio of 
ellipticals with boxy and disky isophotes with luminosity within the observational
uncertainties due to the low number statistics.  In particular,
the inclusion of the contribution of an additional disk component
results in a reduction of the fraction of boxy ellipticals
mainly at low luminosities, while early-type mergers increase the fraction 
of boxy types at high luminosities.  The model predicts a dependence on the 
environment once disk formation by gas accretion is taken into account. 
Galaxy assembly by major mergers occurs in
high density environments at earlier times allowing secular disks to grow
for longer times and hence become a more important component
in terms of total baryonic mass fraction.

We find two populations of disky ellipticals, those formed by
unequal mass mergers, called 1-component disky, and those who grew new disks and 
were originally boxy or 1-component disky ellipticals, called 2-component disky. The 
fraction of 2-component disky ellipticals is a strong function of the luminosity and
environment.  As massive ellipticals form late,
they do not have time to grow secular disks. Disky massive ellipticals therefore are 
mainly 1-component systems, which result from unequal-mass mergers.
The disks in 2-component disky ellipticals will be on average younger than their
bulge populations. We predict that such disks will be mainly found 
in cluster environments where, at luminosities around $M_{B} \sim -18$, half of 
the disky ellipticals should be 2-component systems. If these disks would continue to grow,
they would transform the elliptical into an Sa or Sb galaxy. It is not clear whether
the galaxy at an intermediate stage will look like an S0 galaxy or whether
S0s are sometimes 2-component ellipticals seen under certain projections.
Observations analysing the stellar orbital distribution in disky ellipticals
will be able to distinguish between the two populations of disky ellipticals.

We predict also two populations of boxy elliptical galaxies,
power-law-boxy,  made by 
equal mass mergers of disk galaxies,  and core-boxy, made by any major merger 
but between two elliptical galaxies. We find that the 
fraction of core-to-power-law galaxies increases toward higher luminosities
because of a higher fraction of mergers between bulge dominated 
galaxies. However, we do not find an environmental dependence due to
the self-similar build-up through mergers. The observations of \citet{re01} 
agree very well with our predictions. A larger observational study will be
required to test the relation which we find for a larger magnitude range.

\citet{ben92} argued that ellipticals can be divided in two main categories,
 boxy and disky ellipticals. Our study shows that to recover observational 
trends we need to make assumptions on the formation process of ellipticals 
which introduce additional sub-classes.
The boxy and disky sub-classes have different origins which are 
closely related to the hierarchical build-up of structure and local gas physics.
The two boxy classes differ because of the properties of their progenitors 
which are again dependent on the previous merging history.
The two disky classes differ because of the 
baryonic gas physics, namely the build up of a disk
by gas accretion after the last major merger.
By comparing the predicted fractions of boxy and disky sub-classes with 
observations it will be possible in the future to clarify the importance of 
gas physics coupled to star formation
and the influence of the global dark matter merging history on the origin of early-type
galaxy morphologies.
\newline

The authors would like to thank Ralf Bender, Thorsten Naab, Sukyoung Yi and the referee
 for useful comments. SK acknowledges support by PPARC Theoretical Cosmology Rolling 
Grant and by the STScI where part of this work has been done. AB acknowledges the
support of the Aspen Center of Physics where part of this work was done.



\label{lastpage}

\end{document}